\documentclass[usenatbib]{mn2e}
\usepackage{graphicx,fixltx2e}
\usepackage{wasysym}

\def\rpd{\hbox{rad\,d$^{-1}$}}

\def\chisqr{\hbox{$\chi^2_{\rm r}$}}

\def\sn{\hbox{S/N}}
\def\vrad{\hbox{$v_{\rm rad}$}}
\def\ms{\hbox{m\,s$^{-1}$}}

\def\kms{\hbox{km\,s$^{-1}$}}

\def\vesini{\hbox{$v_{\rm e}\sin i$}}

\def\ptt{\hbox{$10^{-4} I_{\rm c}$}}

\def\degr{\hbox{$^\circ$}}

\def\omeq{\hbox{$\Omega_{\rm eq}$}}
\def\dom{\hbox{$d\Omega$}}

\newcommand{\caii}{\hbox{Ca$\;${\sc II}}}

\begin{document}

\title[Magnetic cycles of the planet-hosting star $\tau$~Bootis: II. a second magnetic polarity reversal]
{Magnetic cycles of the planet-hosting star $\tau$~Bootis: II. a second magnetic polarity reversal
 }

\makeatletter

\def\newauthor{%
  \end{author@tabular}\par
  \begin{author@tabular}[t]{@{}l@{}}}
\makeatother

\author[R.~Fares et al.]{\vspace{1.7mm} 
R.~Fares$^{1,2}$\thanks{E-mail:
rim.fares@ast.obs-mip.fr (RF);
donati@ast.obs-mip.fr (J-FD);
claire.moutou@oamp.fr (CM);  
david.bohlender@nrc-cnrc.gc.ca (DB); 
claude.catala@obspm.fr (CC); 
magali.deleuil@oamp.fr (MD); 
shkolnik@dtm.ciw.edu (ES);,
acc4@st-andrews.ac.uk (ACC);
mmj@st-andrews.ac.uk (MMJ)
gordonwa@uvic.ca (GAHW)},
J.-F.~Donati$^1$, C.~Moutou$^2$, D. Bohlender$^3$, C.~Catala$^4$, M.~Deleuil$^2$ \\ 
\vspace{1.7mm}
{\hspace{-1.5mm}\LARGE\rm
E. Shkolnik$^5$, A.C.~Cameron$^6$, M.M.~Jardine$^6$, G.A.H.~Walker$^7$}\\
$^1$ LATT--UMR 5572, CNRS \& Univ.\ P.~Sabatier, 14 Av.\ E.~Belin, F--31400 Toulouse, France \\
$^2$ LAM--UMR 6110, CNRS \& Univ.\ de Provence,38 rue Fr\'ederic Juliot-Curie, F--13013 Marseille, France \\
$^3$ HIA/NRC, 5071 West Saanich Road, Victoria, BC V9E 2E7, Canada \\ 
$^4$ LESIA--UMR 8109, CNRS \& Univ.\ Paris VII, 5 Place Janssen, F--92195 Meudon Cedex, France \\
$^5$ Department of Terrestrial Magnetism, Carnegie Institution of Washington,  5241 Broad Branch Road, NW, Washington, DC 20015-130 , USA \\ 
$^6$ School of Physics and Astronomy, Univ.\ of St~Andrews, St~Andrews, Scotland KY16 9SS, UK \\
$^7$ 1234 Hewlett Place, Victoria, BC V8S 497, Canada\\
}

\date{2009, MNRAS, submitted}
\maketitle

\begin{abstract}

In this paper, we present new spectropolarimetric observations of the planet-hosting star $\tau$~Bootis, using ESPaDOnS and Narval spectropolarimeters at Canada-France-Hawaii Telescope (CFHT) and Telescope Bernard Lyot (TBL), respectively.

We detected the magnetic field of the star at three epochs in 2008. It is a weak magnetic field of only a few Gauss, oscillating between a predominant toroidal component in January and a dominant poloidal component in June and July. A magnetic polarity reversal was observed relative to the magnetic topology in June 2007. This is the second such reversal observed in two years on this star, suggesting that $\tau$~Boo has a magnetic cycle of about 2 years. This is the first detection of a magnetic cycle for a star other than the Sun. The role of the close-in massive planet in the short activity cycle of the star is questioned.

$\tau$~Boo has strong differential rotation, a common trend for stars with shallow convective envelope. At latitude 40\degr, the surface layer of the star rotates in 3.31 d, equal to the orbital period. Synchronization suggests that the tidal effects induced by the planet may be strong enough to force at least the thin convective envelope into corotation.

$\tau$~Boo shows variability in the \caii\ H \& K and H$\alpha$ throughout the night and on a night to night time scale. We do not detect enhancement in the activity of the star that may be related to the conjunction of the planet. Further data is needed to conclude about the activity enhancement due to the planet.

\end{abstract}


\section{Introduction}
\label{sec:intro}

Hot Jupiters (hereafter HJ) are giant planets orbiting their stars with very short orbital period (less than 10 days), having masses comparable to or larger than that of Jupiter. They represent about 25~\% of all discovered extrasolar planets. Their proximity to their parent stars likely influences their orbital evolution and internal structure; a variety of interactions (generically called Star-Planet interactions or SPI) are also expected to occur. 

Observations suggest that some HJ systems undergo episodes of enhanced variability in their activity (e.g. central emission in \caii\ H \& K or Balmer lines) (\cite{shk03}, \cite{shk05}) or in their mean optical brightness \citep{walker08}. These SPI episodes are apparently phase shifted with orbital conjunctions \citep{shk05}; they are also reported to exhibit longer term fluctuations, suggesting an intermittent ('on-off') nature of SPI \citep{shk08}. 

Two types of interactions were proposed to qualitatively explain the observations: magnetic and tidal interactions \citep{cuntz00}. Magnetic SPI can induce reconnection events between the magnetic field lines of the star and those of the planet; the result is a modulation of the stellar activity with the orbital period. Tidal interactions result in two tidal bulges on the star, and thus, enhanced activity will be modulated by half the orbital period. Potential phase lags (between the orbital conjunction and the epochs of activity variability enhancement) may result from the tilt of the magnetic axis relative to the rotational one \citep{mcivor06}; other scenarios of interaction may also explain this, as adopting the Alfv\'en wing model \citep{preusse06} or considering a non-potential magnetic field configuration for the closed corona of the star \citep{lanza08}. Simulations of \caii\ H \& K light curves of a HJ hosting star also show that the interactions depend on the configuration of the magnetic field which the planet crosses, and thus, disappearance of the SPI may be explained by a change of the stellar magnetic field \citep{cranmer07}.

The magnetic field of the star is therefore expected to play a key role in these interactions, either by triggering them directly (magnetic interactions), or by tracing them indirectly (tidal interactions influencing the generation of the magnetic field, e.g., by enhancing the shear at the base of the convective envelope). Studying the global magnetic field of HJ hosting stars thus appears as a promising tool to investigate quantitatively these SPIs.

With a massive HJ orbiting in $3.31$~d at $0.049$~AU \citep{butler97,Leigh03}, the F7 star $\tau$~Boo is a good candidate for our study. It has a shallow outer convective envelope of about 0.5 $\rm{M_{\jupiter}}$, while the mass of the planet is about 6 $\rm{M_{\jupiter}}$. This star has moderate intrinsic activity \citep{shk05,shk08}, suggesting that it may be a good candidate for detecting any planet-induced activity signatures (of presumably very low amplitude). $\tau$~Boo has a weak magnetic field of few Gauss \citep{catala07,donati08}, that switched its polarity in a year, what makes the study of the magnetic field of this star interesting.

We present in this paper a new multi-epoch spectropolarimetric study of $\tau$~Bootis. The modeling of the large-scale magnetic field of the star, of its differential rotation and its activity  is presented in Sections \ref{sec:mod}, \ref{sec:DR} \& \ref{sec:Activity}. We draw some conclusions in section \ref{sec:dis}.  
 
\section{Observations}
\label{sec:obs}

\indent Spectropolarimetric data of $\tau$~Bootis were collected using ESPaDOnS and Narval. ESPaDOnS is a high resolution spectropolarimeter installed at the 3.6 meter Canada-France-Hawaii Telescope (CFHT) in Hawaii. Narval is a twin instrument, installed at the 2 meter telescope Bernard Lyot (TBL) in France. The spectra obtained using both instruments span the whole optical domain (370 nm to 1048 nm), having a resolution of about 65000. Each spectrum consists of four subexposures taken in different configurations of the polarimeter waveplates, in order to perform a full circular polarization analysis. Data were reduced using a fully automatic reduction tool Libre-Esprit, installed at the CFHT and the TBL. It extracts unpolarized (Stokes I) and circular polarized (Stokes V) spectra of the stellar light. A null spectrum (labeled N) is also produced to confirm that the detected polarization is real and not due to spurious instrumental or reduction effects \citep{Donati97}. 

We collected 67 spectra, on three separate runs in 2008. The first run was on ESPaDOnS from January 19 to January 29, 40 spectra were collected, having good \sn\ ratio (1404/2160 around 700 nm) and permitting a good coverage of the rotation cycle of the star. The two other runs were on Narval, on June (21 - 28) and July (10 - 24). The Stokes V signatures that were detected in June 2008 are small (see Fig. \ref{fig:profils}), the \sn\ ratios around 700 nm vary from 858 to 2118, and from 898 to 1901 for the 19 collected spectra in July 2008 (bad weather condition for 23 July 2008). The complete log is given in Table \ref{tab:logjan} for the ESPaDOnS run, and in Table \ref{tab:logjunjul} for the Narval runs.
\begin{table*}
\caption[]{Journal of January 2008 observations obtained with ESPaDOnS.  Columns 1--8 sequentially list the UT date, the heliocentric Julian date and UT time (both at mid-exposure),
the complete exposure time,  the peak signal to noise ratio (per 2.6~\kms\ velocity bin) of each observation (around 700 nm), the
orbital cycle (using the ephemeris given by Eq.~\ref{eq:eph}), the radial 
velocity (RV) associated with each exposure and the rms noise level (relative to the unpolarized continuum level
$I_{\rm c}$ and per 1.8~\kms\ velocity bin) in the circular polarization profile
produced by Least-Squares Deconvolution (LSD).  }
\begin{tabular}{cccccccc}
\hline
\hline
Date & HJD          & UT      & $t_{\rm exp}$  & \sn & Cycle & \vrad & $\sigma_{\rm LSD}$\\
(2008)    & (2,454,000+) & (h:m:s) & (s) & & (312+) & (\kms)&(\ptt)  \\
\hline
19 Jan  &  486.084820 & 14:01:58 & 4$\times$100 & 1404 & 0.4880 &-16.325 & 0.27 \\
19 Jan  &  486.092120 & 14:12:28 & 4$\times$100 & 1420 & 0.4902 &-16.331 & 0.26 \\
20 Jan  &  487.070470 & 13:41:10 & 4$\times$160 & 1500 & 0.7855 &-16.807 & 0.26 \\
20 Jan  &  487.080380 & 13:55:26 & 4$\times$160 & 1590 & 0.7885 &-16.801 & 0.24 \\
20 Jan  &  487.090870 & 14:10:33 & 4$\times$160 & 1685 & 0.7917 &-16.798 & 0.23 \\
21 Jan  &  488.036410 & 12:52:00 & 4$\times$235 & 1826 & 1.0772 &-16.156 & 0.21 \\
21 Jan  &  488.050970 & 13:12:59 & 4$\times$235 & 1808 & 1.0815 &-16.143 & 0.21 \\
21 Jan  &  488.064370 & 13:32:16 & 4$\times$235 & 1850 & 1.0856 &-16.135 & 0.21 \\
21 Jan  &  488.160300 & 15:50:23 & 4$\times$235 & 1871 & 1.1146 &-16.063 & 0.20 \\
21 Jan  &  488.173630 & 16:09:35 & 4$\times$235 & 1704 & 1.1186 &-16.054 & 0.23 \\
21 Jan  &  488.184770 & 16:25:38 & 4$\times$125 & 1245 & 1.1219 &-16.046 & 0.31 \\
22 Jan  &  489.040310 & 12:57:31 & 4$\times$240 & 1767 & 1.3802 &-16.083 & 0.22 \\
22 Jan  &  489.053420 & 13:16:23 & 4$\times$240 & 1869 & 1.3842 &-16.092 & 0.21 \\
22 Jan  &  489.067010 & 13:35:57 & 4$\times$240 & 1972 & 1.3883 &-16.101 & 0.19 \\
22 Jan  &  489.160330 & 15:50:19 & 4$\times$240 & 2064 & 1.4165 &-16.160 & 0.18 \\
22 Jan  &  489.173410 & 16:09:09 & 4$\times$240 & 2088 & 1.4204 &-16.171 & 0.18 \\
22 Jan  &  489.186580 & 16:28:07 & 4$\times$240 & 1979 & 1.4244 &-16.176 & 0.19 \\
23 Jan  &  490.046950 & 13:06:57 & 4$\times$260 & 1855 & 1.6841 &-16.808 & 0.21 \\
23 Jan  &  490.061030 & 13:27:13 & 4$\times$260 & 1864 & 1.6884 &-16.809 & 0.21 \\
23 Jan  &  490.075200 & 13:47:38 & 4$\times$260 & 1848 & 1.6926 &-16.810 & 0.21 \\
23 Jan  &  490.159790 & 15:49:25 & 4$\times$230 & 1823 & 1.7182 &-16.825 & 0.21 \\
23 Jan  &  490.172450 & 16:07:39 & 4$\times$230 & 1830 & 1.7220 &-16.824 & 0.21 \\
23 Jan  &  490.185120 & 16:25:54 & 4$\times$230 & 1760 & 1.7258 &-16.826 & 0.22 \\
24 Jan  &  491.062630 & 13:29:24 & 4$\times$230 & 2160 & 1.9907 &-16.362 & 0.17 \\
24 Jan  &  491.104600 & 14:29:50 & 4$\times$180 & 1830 & 2.0034 &-16.317 & 0.21 \\
26 Jan  &  493.046930 & 13:06:33 & 4$\times$240 & 1685 & 2.5898 &-16.593 & 0.21 \\
26 Jan  &  493.060120 & 13:25:33 & 4$\times$240 & 1686 & 2.5938 &-16.609 & 0.20 \\
26 Jan  &  493.073290 & 13:44:31 & 4$\times$240 & 1704 & 2.5977 &-16.615 & 0.20 \\
26 Jan  &  493.161890 & 15:52: 5 & 4$\times$220 & 1698 & 2.6245 &-16.708 & 0.22 \\
26 Jan  &  493.174080 & 16:09:38 & 4$\times$220 & 1748 & 2.6282 &-16.724 & 0.21 \\
26 Jan  &  493.186500 & 16:27:31 & 4$\times$220 & 1689 & 2.6319 &-16.728 & 0.22 \\
27 Jan  &  494.046290 & 13:05:31 & 4$\times$240 & 1264 & 2.8915 &-16.677 & 0.26 \\
27 Jan  &  494.059860 & 13:25:03 & 4$\times$240 & 1410 & 2.8956 &-16.657 & 0.22 \\
27 Jan  &  494.072960 & 13:43:55 & 4$\times$240 & 1367 & 2.8995 &-16.650 & 0.23 \\
27 Jan  &  494.160790 & 15:50:23 & 4$\times$220 & 1502 & 2.9260 &-16.579 & 0.24 \\
27 Jan  &  494.173120 & 16:08:09 & 4$\times$220 & 1430 & 2.9298 &-16.579 & 0.26 \\
27 Jan  &  494.185340 & 16:25:44 & 4$\times$220 & 1573 & 2.9335 &-16.569 & 0.25 \\
29 Jan  &  496.118430 & 14:49:09 & 4$\times$300 & 1461 & 3.5170 &-16.412 & 0.23 \\
29 Jan  &  496.134930 & 15:12:54 & 4$\times$300 & 1426 & 3.5220 &-16.420 & 0.23 \\
29 Jan  &  496.150890 & 15:35:54 & 4$\times$300 & 1847 & 3.5268 &-16.452 & 0.19 \\

\hline
\end{tabular}
\label{tab:logjan}
\end{table*}
\begin{table*}
\caption[]{Journal of June and July observations obtained with Narval.  Columns 1--8 indicate the same parameters as table \ref{tab:logjan}.
  }    
\begin{tabular}{cccccccc}
\hline
\hline
Date  & HJD          & UT      & $t_{\rm exp}$  & \sn & Cycle & \vrad & $\sigma_{\rm LSD}$ \\
(2008)    & (2,454,000+) & (h:m:s) & (s) & & (+358) & (\kms)&(\ptt)  \\
\hline
21 June  &  639.358680 & 20:33:14 & 4$\times$300 & 858 & 0.7600 &-16.868 & 0.45 \\
21 June  &  639.381310 & 21: 5:49 & 4$\times$600 & 1200 & 0.7669 &-16.843 & 0.32 \\
21 June  &  639.411470 & 21:49:16 & 4$\times$600 & 1359 & 0.7760 &-16.832 & 0.28 \\
22 June  &  640.362370 & 20:38:39 & 4$\times$300 & 1471 & 1.0630 &-16.179 & 0.24 \\
22 June  &  640.384990 & 21:11:14 & 4$\times$600 & 1997 & 1.0699 &-16.154 & 0.18 \\
22 June  &  640.415150 & 21:54:39 & 4$\times$600 & 2118 & 1.0790 &-16.132 & 0.17 \\
25 June  &  643.375010 & 20:57:10 & 4$\times$600 & 1660 & 1.9725 &-16.397 & 0.22 \\
27 June  &  645.358100 & 20:33: 2 & 4$\times$300 & 1374 & 2.5712 &-16.538 & 0.25 \\
28 June  &  646.351660 & 20:23:53 & 4$\times$300 & 1363 & 2.8712 &-16.704 & 0.25 \\
\hline
10 July  &  658.351740 & 20:25:20 & 4$\times$600 & 1873 & 6.4939 &-16.368 & 0.19 \\
14 July  &  662.352640 & 20:27: 6 & 4$\times$600 & 1355 & 7.7017 &-16.914 & 0.28 \\
15 July  &  663.348890 & 20:21:49 & 4$\times$600 & 1810 & 8.0025 &-16.432 & 0.20 \\
17 July  &  665.352950 & 20:27:54 & 4$\times$450 & 1880 & 8.6075 &-16.666 & 0.20 \\
17 July  &  665.376170 & 21: 1:20 & 4$\times$450 & 1843 & 8.6145 &-16.673 & 0.20 \\
18 July  &  666.347980 & 20:20:52 & 4$\times$450 & 1901 & 8.9079 &-16.635 & 0.19 \\
18 July  &  666.371210 & 20:54:18 & 4$\times$450 & 1854 & 8.9149 &-16.585 & 0.19 \\
19 July  &  667.374180 & 20:58:42 & 4$\times$450 & 1492 & 9.2177 &-15.925 & 0.26 \\
19 July  &  667.397390 & 21:32: 8 & 4$\times$450 & 1196 & 9.2247 &-15.912 & 0.33 \\
20 July  &  668.373790 & 20:58:15 & 4$\times$450 & 1050 & 9.5194 &-16.390 & 0.36 \\
20 July  &  668.397020 & 21:31:43 & 4$\times$450 & 1229 & 9.5265 &-16.427 & 0.32 \\
21 July  &  669.349580 & 20:23:31 & 4$\times$450 & 1172 & 9.8140 &-16.843 & 0.32 \\
21 July  &  669.372800 & 20:56:57 & 4$\times$450 & 1259 & 9.8210 &-16.814 & 0.30 \\
22 July  &  670.348550 & 20:22: 9 & 4$\times$450 & 1846 & 10.1156 &-16.081 & 0.19 \\
22 July  &  670.371770 & 20:55:35 & 4$\times$450 & 1791 & 10.1226 &-16.047 & 0.20 \\
23 July  &  671.348660 & 20:22:25 & 4$\times$450 & 898 & 10.4175 &-16.156 & 0.49 \\
23 July  &  671.371890 & 20:55:53 & 4$\times$450 & 571 & 10.4245 &-16.141 & 1.03 \\
24 July  &  672.352850 & 20:28:34 & 4$\times$450 & 1717 & 10.7207 &-16.832 & 0.21 \\
24 July  &  672.376080 & 21: 2: 2 & 4$\times$450 & 1649 & 10.7277 &-16.825 & 0.23 \\
\hline
\end{tabular}
\label{tab:logjunjul}
\end{table*}

All data are phased with the same orbital ephemeris as that of \citet{catala07} and \citet{donati08}:
\begin{equation}
T_0 = \mbox{HJD~}2,453,450.984 + 3.31245 E
\label{eq:eph}
\end{equation}
with phase 0.0~indicating the first conjunction (i.e. with the planet farthest from the observer). 

The Zeeman signatures of $\tau$~Boo are extremely small. We use Least Squares Deconvolution (LSD) to improve the S/N ratio of our data. This technique consists of deconvolving the observed spectra using a line mask. The line mask was computed using a Kurucz model atmosphere with solar abundances; effective temperature and logarithmic gravity (in \hbox{cm\,s$^{-2}$}) are set to $6250$~K and $4.0$, respectively. The line mask includes the moderate to strong lines present in the optical domain (those featuring central depths larger than 40~\% of the local continuum, before any macro turbulent or rotational broadening, about 4,000 lines throughout the whole spectral range) but excludes the strongest, broadest features, such as Balmer lines, whose Zeeman signature is strongly smeared out compared to those of narrow lines. The typical multiplex gain for the polarization profiles is between 25 and 30, implying noise levels in LSD polarization profiles as low as 20 parts per million (ppm).

Radial velocities of the star can be obtained by fitting the Stokes I profiles (unpolarized profiles) with a Gaussian to each profile. The values we obtain are in good agreement with the expectations (Fig. \ref{fig:vrad}), when using the orbital ephemeris of \citet{catala07}. All spectra are automatically corrected from spectral shifts resulting from instrumental effects (e.g. mechanical flexures, temperature or pressure variations) using telluric lines as a reference. Though not perfect, this procedure allows spectra to be secured with radial velocity (RV) precision of better than 30~\ms~\citep{moutou07,morin08}. All spectra are corrected from the orbital motion.
\begin{figure}
\includegraphics[height=.25\textheight]{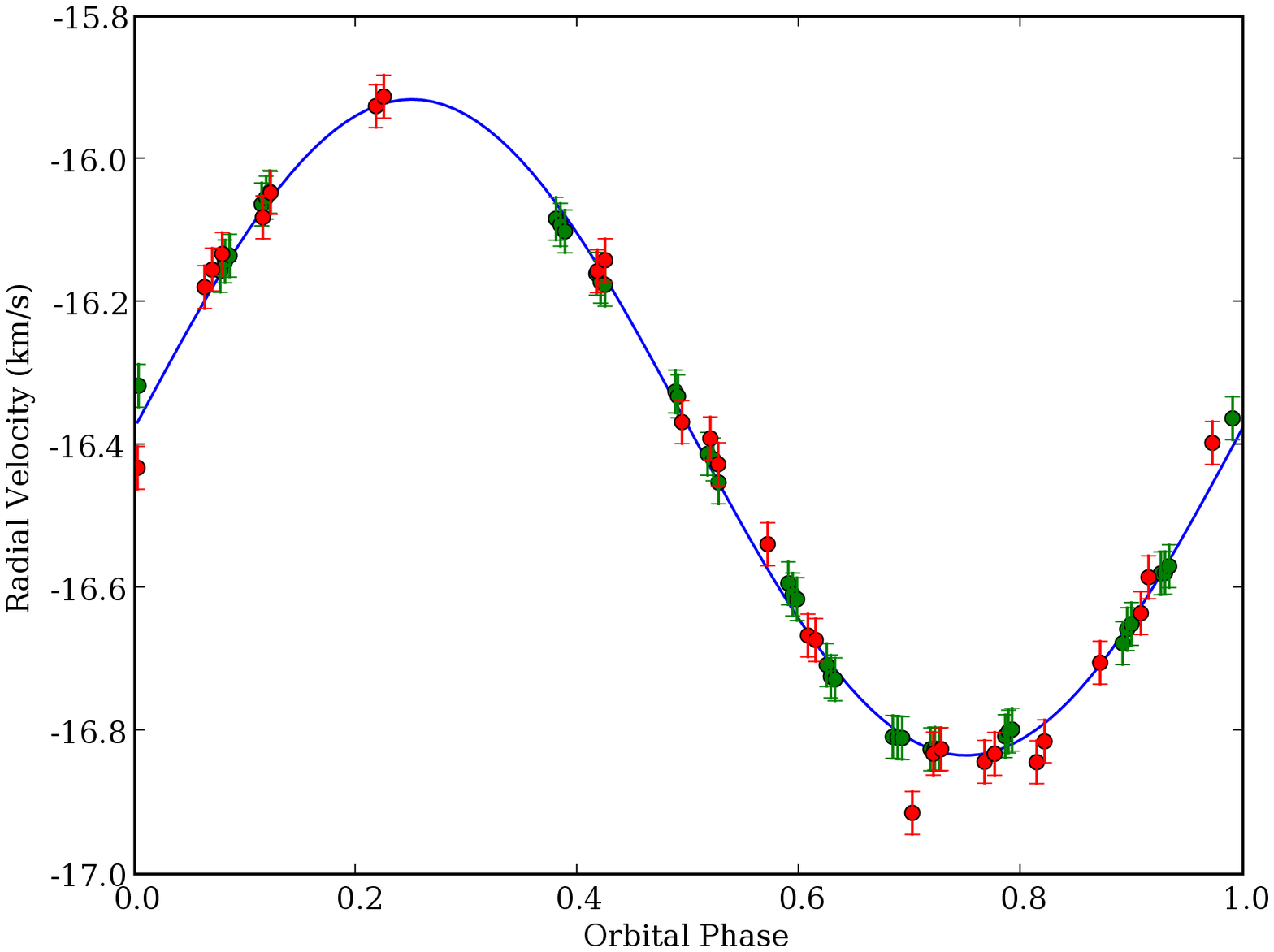}
\caption[]{Radial velocities of $\tau$~Boo derived from our Narval (red dots) and ESPaDOnS (green dots) spectra as a function of orbital phase, with their error bars ($20 - 30$~\ms). The radial velocity model plotted here (blue full line) corresponds to our fit of the data, giving orbital solution compatible with those in the literature \citep{butler06}. }
\label{fig:vrad}
\end{figure}

\section{Magnetic modelling}
\label{sec:mod}

\subsection{Model description}

We use Zeeman Doppler Imaging (hereafter ZDI) to get magnetic maps of $\tau$~Boo and an estimation of its differential rotation. ZDI is a tomographic imaging technique; it consists of inverting series of Stokes V profiles into a map of the parent magnetic topology, i.e. both the location of magnetic spots and the orientation of the field lines. We use the most recent version of the code \citep{donati06}, which  describes the field by its poloidal and toroidal components, both expressed in terms of spherical harmonic expansions. This has the advantage that both simple and complex magnetic topologies can be reconstructed. The energy of axisymmetric ($l < m/2$) and non-axisymmetric modes is easily calculated from the coefficients of the spherical harmonics.

The surface of the star is decomposed into 50000 small cells of similar area. The procedure takes into account the contribution of each cell to the reconstructed profile. This is done by iteratively comparing the synthetic profile to the observed one, until they match within the error bars (i.e. reduced chi-square $\chisqr \sim 1$). Since the inversion problem is ill-posed, we use Maximum Entropy as a criterion that ensures a unique solution.

The models we use for computing the local Stokes I and V profiles associated with each grid cell are quite simple. Stokes I is modelled by a Gaussian with a FWHM of 11~\kms, Stokes V is modelled assuming the weak field approximation, i.e. $V \propto g B_{los} dI/dv$~where $B_{los}$~is  the line-of-sight projected magnetic field at a selected point at the surface of the star and $g$ is the mean land\'e factor (set to 1.2). The magnetic field detected on $\tau$~Boo (5-10~G, \cite{catala07}, \cite{donati08}) demonstrates that this approximation is valid.

\subsection{Magnetic maps}
\label{sec:MM}

Given the projected equatorial velocity of $\tau$~Boo ($\vesini = 15.9$~\kms) and the FWHM of the local profile intensity of 11~\kms, there are $\sim$~9
 spatial resolution elements around the equator. In practice, the information added to the maps when considering a degree of spherical harmonics greater than 8 is of little significance; we therefore use $l_{max}=8$.
As in \citet{catala07}, the inclination angle of the rotation axis of the star with respect to the line-of-sight is assumed to be $i =40\degr$.
Fitting the N (null) profiles (instead of Stokes V profiles) assuming no magnetic field at the surface of the star gives a \chisqr~level varying between 0.85 and 0.9 depending on the epoch (indicating that our error bars are slightly overestimated by about 5~\%). We therefore fit all our Stokes V data to a level of $\chisqr=0.9$.
\begin{figure*}
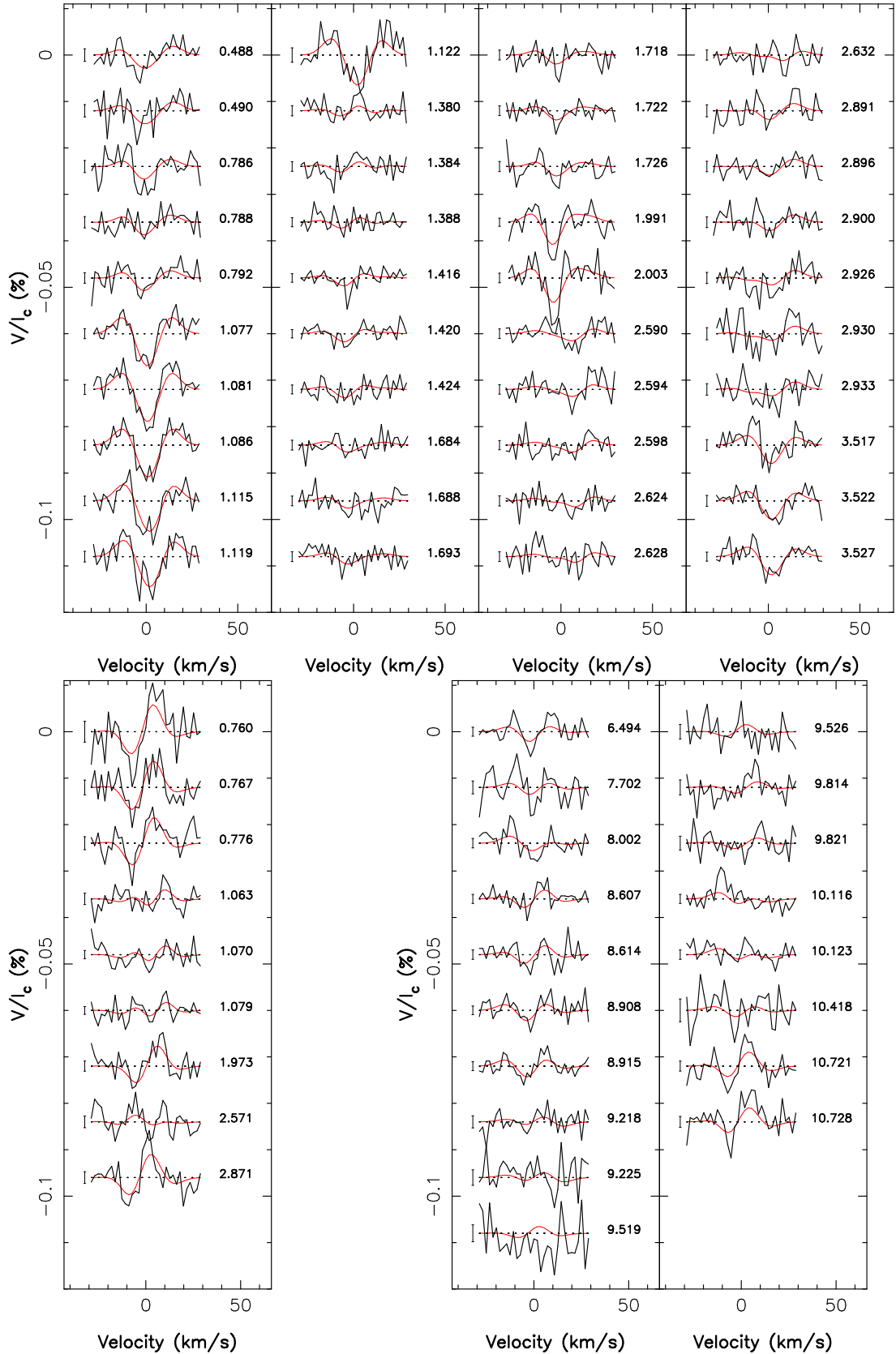

\center{\hbox{\hspace{20mm}\includegraphics[scale=0.6,angle=-90]{SpectreJan08.ps}}
        \hbox{\hspace{20mm}\includegraphics[scale=0.6,angle=-90]{SpectreJun08.ps}\hspace{20mm}\includegraphics[scale=0.6,angle=-90]{SpectreJul08.ps}}}
\caption[]{Maximum-entropy fits (thin red line) to the observed Stokes $V$~LSD 
profiles (thick black line) of $\tau$~Boo for 2008 January (top), 2008 June (bottom, left) and 2008 July (bottom, right) are displayed.  The orbital cycle of each observation 
(as listed in Tables \ref{tab:logjan} \& \ref{tab:logjunjul}) and 1$\sigma$ error bars are also shown next to each profile.  }
\label{fig:profils}
\end{figure*}

\begin{figure*}
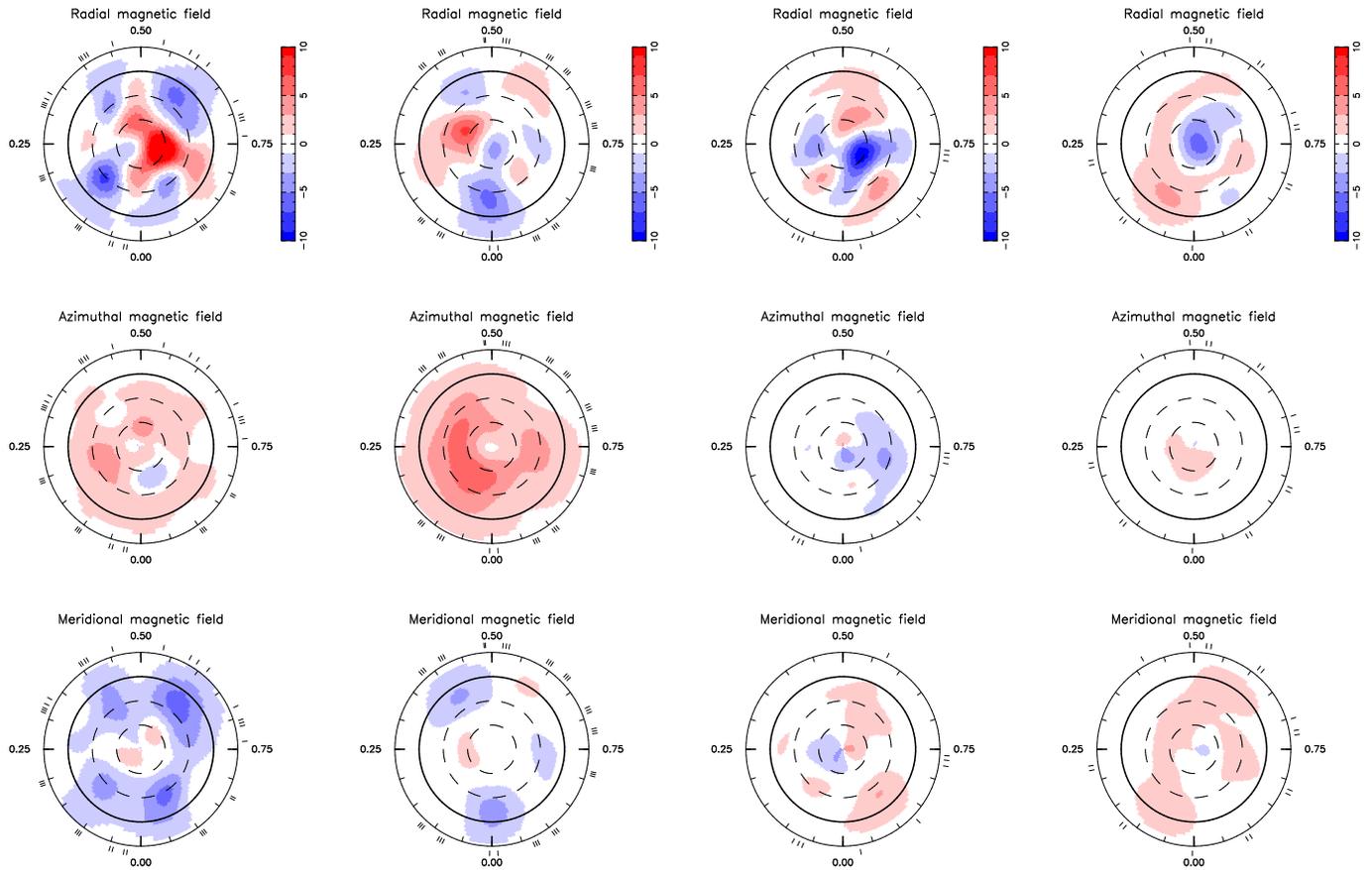

\center{\hbox{\includegraphics[scale=0.4]{mapJun07.ps}\hspace{6mm}
              \includegraphics[scale=0.4]{mapJan08.ps}\hspace{6mm}
              \includegraphics[scale=0.4]{mapJun08.ps}\hspace{6mm}
	      \includegraphics[scale=0.4]{mapJul08.ps}}} 
\caption[]{Maximum-entropy reconstructions of the large-scale magnetic topology of
$\tau$~Boo as derived from our 2008 January (second column), 2008 June (third column) and 2008 July (right column) data sets are displayed. The radial, azimuthal and meridional components of the field (with magnetic flux values labelled in G).
The star is shown in flattened polar projection down to latitudes of $-30\degr$, with the equator depicted as a bold circle and parallels as dashed circles. Radial ticks around each plot indicate orbital phases of observations. The June 2007 magnetic map of $\tau$~Boo \citep{donati08} is added in the left column to ease comparison. }
\label{fig:maps}
\end{figure*}

\subsubsection{January 2008}

The observed and reconstructed Stokes V profiles of January 2008 are shown in Fig. \ref{fig:profils} (top panel). The signatures vary strongly in shape and amplitudes over rotational phases (e.g. signatures of January 20 compared to those of January 22), which suggests a complex magnetic topology. Reconstructed profiles fit the observed ones within the error bars. The corresponding magnetic map is shown in Fig. \ref{fig:maps} (second column). 

$\tau$~Boo has a weak surface magnetic field of just a few Gauss, reaching a maximum strength at the surface of the star of about 5 to 10 G. The toroidal field dominates (contributing to 62~\% of the total magnetic energy) in the form of a ring of mainly axisymmetric azimuthal field encircling the whole star. The radial poloidal field shows a more complex topology, with two regions of negative polarity at intermediate latitudes, and a positive one. None of those magnetic regions dominates the others, contrary to June 2007 (when the positive polarity contributes more than the negative polarity). Axisymmetric modes enclose 20~\% of the poloidal field energy (see Table \ref{tab:field}). 

\subsubsection{June 2008}

Fig. \ref{fig:profils} and \ref{fig:maps} show the reconstructed Stokes V profiles and map of $\tau$~Boo for June 2008 data (left bottom panel in Fig. \ref{fig:profils}~and third column in Fig.~\ref{fig:maps}). 

Phase coverage is poor. Unlike in January 2008, the dominant component of the field is the poloidal one, enclosing 87~\% of the total magnetic energy: the reconstructed map shows, for the radial field, three regions of positive polarity at low and intermediate latitudes and two dominant regions of negative polarity at intermediate and high latitudes.  The absence of any reconstructed feature in the interval of phases 0.10 to 0.55 is an effect of the poor phase coverage at this epoch. For the poloidal field, axisymmetric and non-axisymmetric modes have almost the same contribution to the poloidal energy ($\sim33$~\%), while axisymmetric modes dominate the toroidal field (enclosing 68~\% of the toroidal field energy).

\subsubsection{July 2008}

The reconstructed profiles for July 2008 fit the observations at $\chisqr$~of 0.9 (see the right bottom panel of Fig. \ref{fig:profils}). The data covers four rotation periods, and are well sampled over the rotational cycle. The reconstructed map shows a field of a few Gauss, mainly poloidal (91~\% of the energy), with a weak toroidal component (right column of Fig. \ref{fig:maps}). The radial field shows a net negative polarity region around the pole, another one at low latitudes, and three positive regions at intermediate latitudes.

\section{Differential rotation}
\label{sec:DR}
\subsection{Method}

To get an estimation of the differential rotation, we use the method described by \citet{donati03} and \citet{morin07}. 
We first consider that the rotation at the surface of the star follows :
\begin{eqnarray}
\Omega(\theta) = \omeq - \dom  \sin^2\theta
\label{eq:DR}
\end{eqnarray} 
where $\Omega(\theta)$ and $\omeq$ are respectively the angular velocities at a latitude $\theta$~and at the equator, and $\dom$~is the difference in rotation rate between the pole and the equator. When the star is differentially rotating, magnetic regions will produce spectral signatures that do not repeat identically from one rotation to another. Measuring the recurrence rate from magnetic regions located at various latitudes gives access to the amount of surface shear. For each pair of (\omeq, \dom), we reconstruct a magnetic image at a given information content. We then choose the pair of parameters that produces the best fit to the data and thus gives the smallest $\chisqr$. In practice, when differential rotation is detected in the data, values of $\chisqr$ for all the reconstructed maps form a paraboloid, whose fit yields the optimum differential rotation parameters. \citet{donati08} applied this method on $\tau$~Boo and found $\omeq = 2.10 \pm 0.04$~\rpd~and $\dom = 0.50 \pm 0.12$~\rpd~as parameters of its differential rotation.

\subsection{Results}

For January 2008, our data covered about three rotations. The differential rotation parameters we find are $\omeq = 1.86 \pm 0.02$~\rpd and $\dom = -0.18 \pm 0.07$~\rpd, implying apparently an anti-solar differential rotation (very scarcely or never observed in F stars \citep{reiners07}). The field for January 2008 is mainly toroidal, i.e. the component providing the strongest contribution to the energy is the azimuthal one; being mostly axisymmetric, this component does not carry much information about the differential rotation. We therefore decided to estimate the parameters of differential rotation using the radial field map only. The parameters we obtain for this second fit are $\omeq=1.93 \pm 0.02$~\rpd and $\dom=0.28 \pm 0.10$~\rpd, more compatible with the results of \citet{donati08}.

For June and July 2008, the paraboloids are well defined, giving approximately the same parameters for the differential rotation. $\omeq = 2.05 \pm 0.04$~\rpd and $\dom = 0.42 \pm0.10 $~\rpd for June 2008, and $\omeq = 2.12 \pm 0.12$~\rpd and $\dom = 0.5 \pm 0.15$~\rpd for July 2008. It implies that $\tau$~Boo rotates in 3 d at the equator while in 3.9 d at the poles. The orbital period being 3.31 d, we infer that latitude $\sim 38\degr$ is rotating synchronously with the planet orbital motion, in agreement with the results of \citet{donati08}.

\section{Activity indicators}
\label{sec:Activity}

We also studied the variation of the residual emission of various activity proxies (in particular, H$\alpha$, \caii\ H \& K). This is done for our three runs by calculating a mean profile (per spectral line) for all the spectra, subtracting it from each spectrum, and then calculating the equivalent width of the residual emission by fitting this residual with a Gaussian profile (we find that this procedure minimizes the error bar on the estimated flux). The signatures of the residual emission for the \caii\ H \& K are extremely small (smaller than 0.5~\% of the unpolarized continuum). Thanks to the better quality of the spectra around 700 nm, H$\alpha$~shows more accurate signatures. The activity variations show similar trends in \caii\ H \& K and H$\alpha$~(see Fig. \ref{fig:halphafit}), a clear positive correlation is observed between both activity indicators.

\begin{figure*}
\center{\hbox{\includegraphics[scale=0.4]{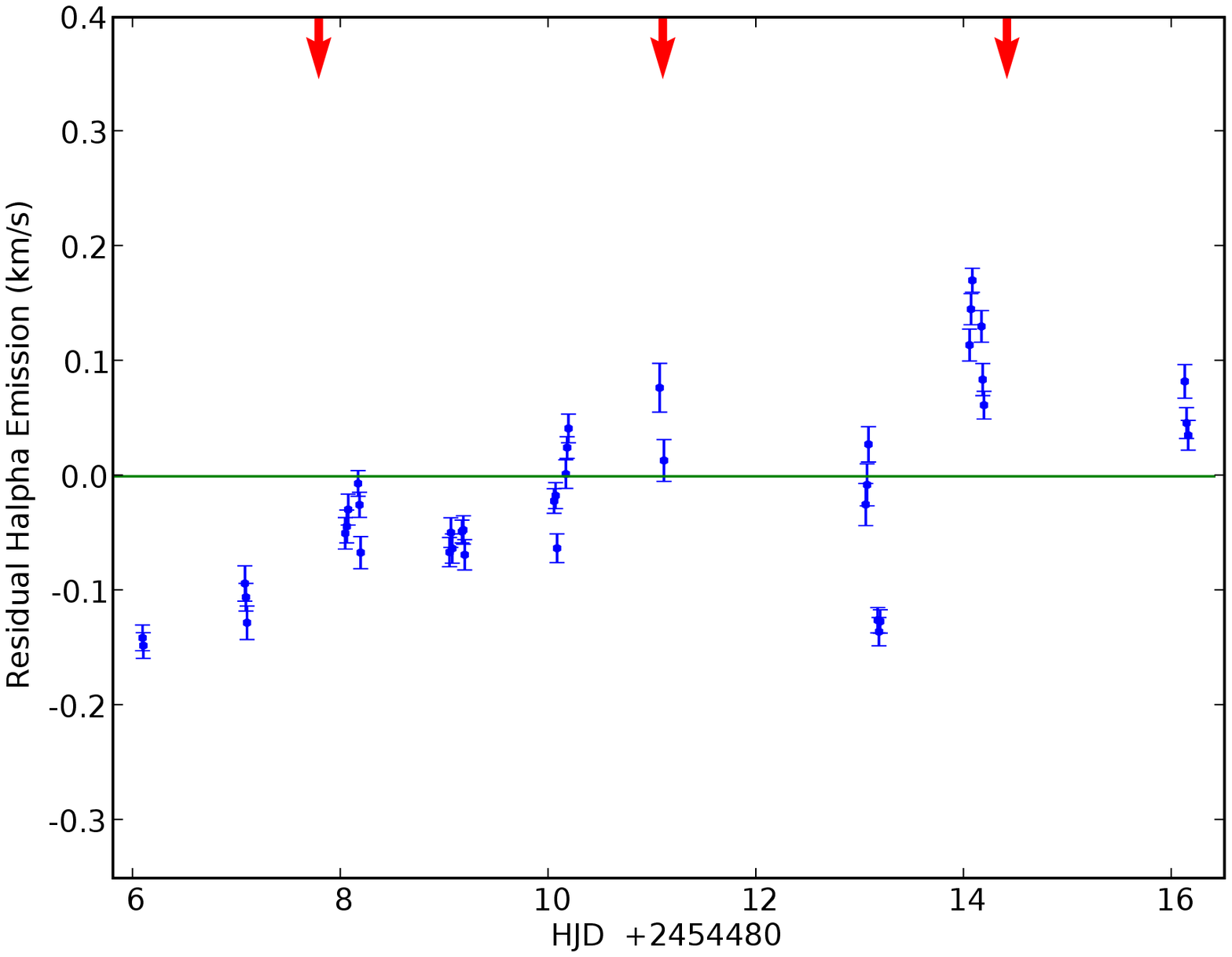}\hspace{7mm}
              \includegraphics[scale=0.4]{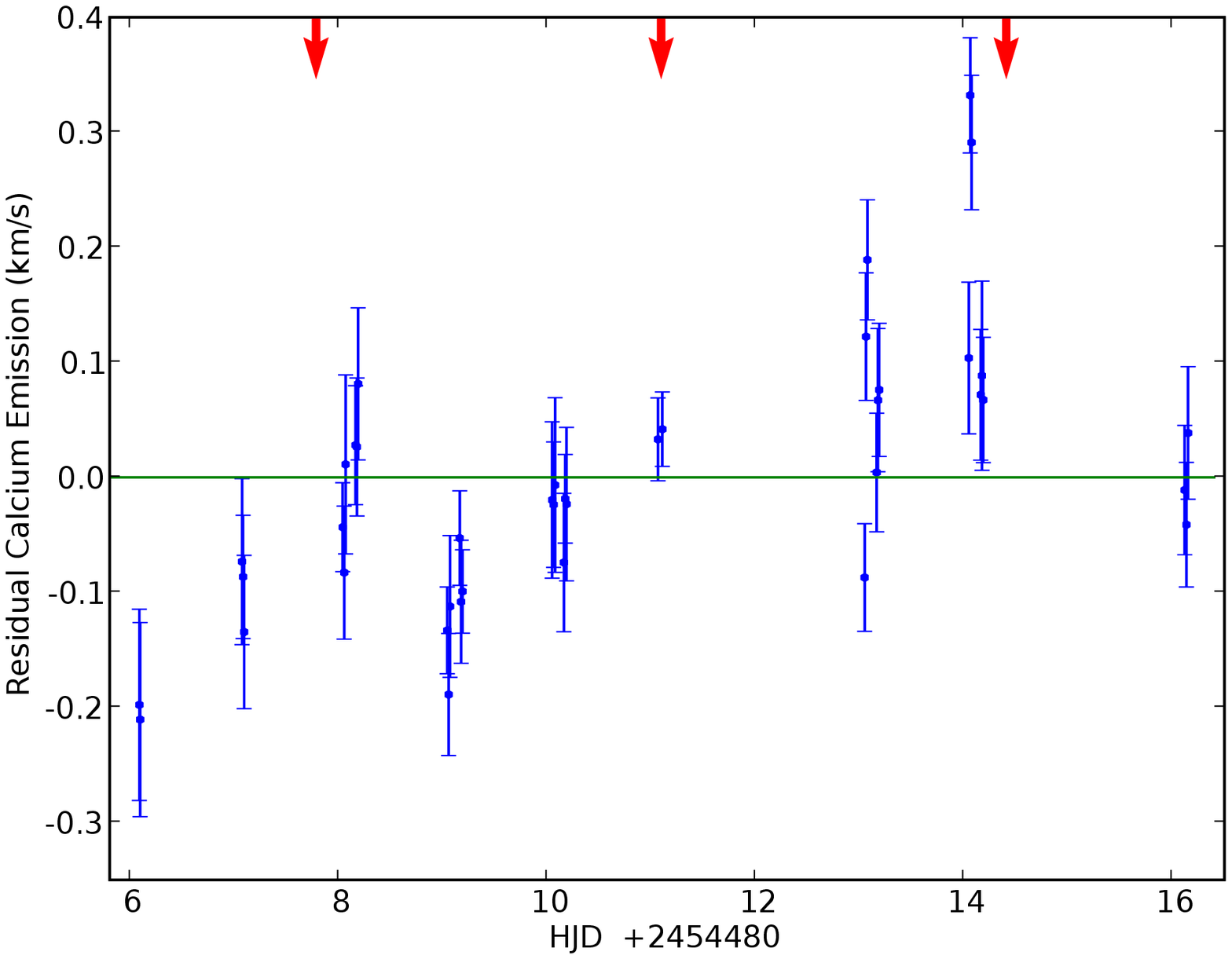}}}
\caption[]{Residual signature of H$\alpha$~(left panel) and Calcium (right panel) as a function of the Heliocentric Julian Date for the ESPaDOnS run January 2008. Red arrows mark the time of opposition. The activity variations show similar trends in \caii\ H \& K and H$\alpha$.}
\label{fig:halphafit}
\end{figure*}

$\tau$~Boo shows intrinsic variability during each night and a night-to-night variability. The clearest trend of January data is a long-term evolution over 10 days unrelated to either the rotational period or the orbital period. To investigate the period on which the activity varies, we fitted the activity residuals using a single sine wave. The fit for all the observed epochs is poor. A small enhancement is observed around the orbital phase 0.8, but it is not significant because of the high intrinsic variability of the star. This enhancement is no longer visible once the long term trend is removed.

\section{Discussion and conclusion}
\label{sec:dis}

We observed $\tau$~Boo during three epochs in 2008: January, June and July and reconstructed the corresponding magnetic maps. A weak magnetic field (maximum intensity 5~-~10~Gauss) is present at the surface of the star, whose configuration varies from epoch to epoch (i.e. on a time scale of months). As for June 2006 \citep{catala07} and June 2007 \citep{donati08}, the field shows a dominant poloidal component in June and July 2008, but it is predominantly toroidal in January 2008. Table \ref{tab:field} lists the main field properties at each epoch. 
\begin{table*}
\caption[]{Average magnetic field on the surface of the star, percentage of the toroidal energy relative to the total one, percentage of the energy contained in the axisymmetric modes of the poloidal component, and percentage of the energy contained in the modes of $l\leq2$ of the poloidal component for each epoch of observation. Data for June 2007 are taken from \cite{donati08}. }    
\begin{tabular}{ccccc}
\hline
& B (G) & \% toroidal & \% axisymmetric modes in poloidal&\% $l\leq2$~in poloidal\\
\hline
\hline
June 2007 & 3.7 & 17 & 60 & 52\\
January 2008 & 3.1 & 62 & 20 &50\\
June 2008 & 2.3 & 13 & 36& 36 \\ 
July 2008 & 1.7 & 9 & 62& 47\\
\hline
\end{tabular}
\label{tab:field}
\end{table*}

Our observations reveal a new global magnetic polarity reversal between June 2007 and June 2008 (following the one reported in \citet{donati08} that occurred between June 2006 and June 2007). This switch is observed for the three components of the field. In June 2008, radial field shows positive magnetic regions near the pole and negative ones at lower latitudes, while the opposite is observed for June 2007 \citep{donati08}. The change in polarity is also observed for both azimuthal and meridional fields. The July 2008 map confirms this conclusion.

Our observation in January 2008 shows that the field has evolved compared to June 2007, but features no global polarity switch. This suggests that the magnetic cycle of $\tau$~Boo is about 2 years, much shorter than that of the Sun ($\sim 22$ years). Comparing June 2007 and January 2008 maps, the azimuthal field becomes stronger in the latter and the radial field weaker, showing a dephasing between the poloidal and toroidal fields, as for the Sun \citep{charbonneau05}.

To carry out a slightly more quantitative analysis, we calculated the (signed) magnetic flux for the radial and azimuthal components on the Northern hemisphere of the star. In the particular case of $B_{r}$, the magnetic flux is counted positive for latitudes higher than 30\degr~and negative for latitudes between 0\degr~and 30\degr~to take into account the contribution of both dipolar and quadrupolar terms of the poloidal field. We fitted simultaneously the two fluxes with sine waves of equal period, the period being varied over a range of 100 - 900 d. We find that the best fit is obtained at 800~d (2.2~years) (Fig. \ref{fig:flux}) and at 250~d (8~months); the 800~d minimum is broader and potentially more likely. This analysis also suggests that the toroidal field is shifted (by 18~\% of the period of 250~d and 28~\% of the period of 800~d) with respect to the poloidal field. A phase shift between the poloidal and toroidal components is also observed for the Sun (of 25~\% of the cycle length, \cite{charbonneau97,jouve07}).
We can safely conclude that the cycle of $\tau$~Boo is at least 10 times shorter than that of the Sun. We caution that this result needs confirmation from additional data, e.g. obtaining a denser monitoring of the magnetic topology over several successive cycles.

\begin{figure}
\includegraphics[height=.25\textheight]{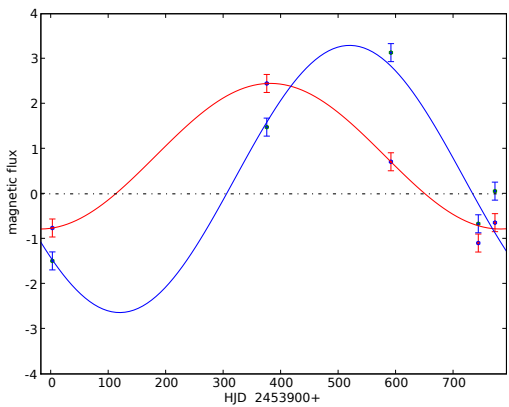}
\caption[]{Fluxes of the radial field (red) and azimuthal field (blue) vs HJD, calculated for the Northern hemisphere of the star. In the particular case of $B_{r}$, the magnetic flux is counted positive for latitudes higher than 30\degr~and negative for latitudes between 0\degr~and 30\degr~to take into account the contribution of both dipolar and quadrupolar terms of the poloidal field. The best sinusoidal fit for $P=800$~d is plotted (see text for more details), using the same colors. }
\label{fig:flux}
\end{figure}

We measured the differential rotation of the star for our three observed epochs. $\tau$~Boo has a strong one, the latitudinal angular rotation shear being $\dom = 0.46 \pm0.08 $~\rpd. Our results are in good agreement with \cite{donati08}. \cite{catala07} and \cite{reiners06} found also similar results, using the Fourier Transform Method described by \cite{reiners02}. $\tau$~Boo's equator rotates in 3 d, while its pole in 3.9 d. This means that a latitude at $\sim40 \degr$ is synchronized with the planet. In contrast to the magnetic field, the differential rotation of $\tau$~Boo has not changed over two years of observations. This is similar to what is observed on the Sun. The average differential rotation calculated from 4 runs is $\dom = 0.43 \pm 0.12$~\rpd. We note that in January 2008 run, the differential rotation measured from the radial component only is slightly weaker than (though still compatible with) the average differential rotation.

We find that activity signatures in usual spectral indexes are very weak (0.5 per cent of the unpolarized continuum). The star shows variability during the night, and from one night to another. The night-to-night variability shows the same trend in \caii\ H \& K and H$\alpha$. We do not see a clear correlation of the residual emission with any sensible period. Our data does not allow us to detect potential activity enhancement due to the planet; such enhancement (if any) is smaller than the short and long term intrinsic variability that we see in the spectra.

$\tau$~Boo is nevertheless a good candidate to study SPI. This is the first star where a magnetic cycle is observed. The generation of the magnetic field of F-type stars is due to dynamo mechanisms, thought to be operating at the base of the convective envelope (the tachocline). $\tau$~Boo's observed magnetic cycle is accelerated compared to that of the Sun. Is it a result of the very shallow convection of $\tau$~Boo ($M_{\rm conv} = 0.5~M_{\jupiter}$)? To answer this question, one has to compare this result to other studies on similar stars (having shallow convective envelope). \citet{marsden06} and \citet{jeffers08} studied the magnetic field of the G0 star HD 171488, which also feature a shallow convective envelope and a strong level of differential rotation. They however did not notice any global changes in its magnetic field over two years; no magnetic polarity switch is observed over four years (Donati, private communication) which means that its magnetic cycle is longer than eight years. One difference between $\tau$~Boo and HD 171488 is that the former is orbited by a HJ. The presence of the HJ at small orbital distance may be responsible for the accelerated cycle of $\tau$~Boo, by synchronizing the outer convective envelope of the star (due to tidal interactions) and enhancing the shear at the tachocline. Such a scenario should however be studied in details. The synchronicity of the outer envelope and its decoupling with the stellar interior are still not well understood. Simulations of the effect of a perturbation due to the planet in the dynamo generation will be attempted; observing more F stars, with and without HJ, and studying their magnetic cycles would permit us to give constraints to the models and draw conclusions on the effect of the planet. 
\section*{Acknowledgments}
This work is based on observations obtained with ESPaDOnS at the Canada-France-Hawaii Telescope (CFHT) and with NARVAL at the T\'elescope Bernard Lyot (TBL). CFHT/ESPaDOnS are operated by the National Research Council of Canada, the Institut National des Sciences de l'Univers of the Centre National de la Recherche Scientifique (INSU/CNRS) of France, and the University of Hawaii, while TBL/NARVAL 
are operated by INSU/CNRS. We thank J. Morin, V. Petit and G. Wade for collecting some of the ESPaDOnS data for us, and the CFHT and TBL staff for their help during the observations. We also thank the referee, A. Lanza, for valuable comments on the manuscript.

\bibliography{paper}
\bibliographystyle{mn2e}

\end{document}